**Optical properties of Xe color centers in diamond**

Russell Sandstrom[1,*], Li Ke[2,*], Aiden Martin[3], Ziyu Wang[2], Mehran Kianinia[1], Ben Green[4], Wei-bo Gao[2,#], Igor Aharonovich[1,#]

1. School of Mathematical and Physical Sciences, University of Technology Sydney, Ultimo, New South Wales 2007, Australia.
2. Division of Physics and Applied Physics, School of Physical and Mathematical Sciences, Nanyang Technological University, 21 Nanyang Link, Singapore 637371, Singapore.
3. Lawrence Livermore National Laboratory, Livermore, California 94550, USA
4. Department of Physics, University of Warwick, Coventry, CV4 7AL, UK

*contributed equally to this work*

#corresponding authors: igor.aharonovich@uts.edu.au and wbgao@ntu.edu.sg

*Optical properties of color centers in diamond have been the subject of intense research due to their promising applications in quantum photonics. In this work we study the optical properties of Xe related color centers implanted into nitrogen rich (type IIA) and an ultrapure, electronic grade diamond. The Xe defect has two zero phonon lines at ~ 794 and 811 nm, which can be effectively excited using both green and red excitation, however, its emission in the nitrogen rich diamond is brighter. Near resonant excitation is performed at cryogenic temperatures and luminescence is probed under strong magnetic field. Our results are important towards the understanding of the Xe related defect and other near infrared color centers in diamond.*

## I.  Introduction

Defects in diamond, (also known as color centers, are important for a variety of applications spanning quantum nanophotonics, bio labelling, light emitting diodes, superconductivity and spintronics[1-6]. Some of these defects – namely the nitrogen vacancy (NV)[7], the silicon vacancy (SiV)[8-11] and more recently the germanium vacancy (GeV)[12,13] have been subject to rigorous research owing to their relatively known crystallographic and electronic structures. Now, however, there is an increased interest in the study of color centers emitting further in the infra-red.

An example of near infra-red color centers in diamond is the Xe related defect with two zero phonon lines (ZPLs) at 794 nm and at 812 nm[14-16]. The crystallographic structure of the Xe color center is still under debate, however it is believed to be an interstitial xenon splitting two vacancies along the <111> crystallographic direction[17]. This was supported by polarized luminescence measurements, which assign the high and low energy ZPLs respectively to σ-σ (XY-XY for absorption and emission, respectively) and σ-π (XY-Z) transitions at a <111>-oriented center[16]. A particularly promising aspect of this color center is the demonstrated ability to generate it via ion implantation and a relatively low Huang Rhys factor of ~ 0.3, indicating that most of the photons are emitted into the ZPL.

In this work, we study in detail the optical properties of the Xe defect, and show that its intensity is strongly dependent on the host material and nitrogen concentration. A series of variable dose Xe ion implantations is performed to determine the emission brightness. Furthermore, the lifetime and polarization properties of the Xe related defect are measured and photoluminescence (PL) excitation at cryogenic temperatures is performed. Our results show promise for using Xe related sources in spectroscopy and nanophotonics applications.

## II. EXPERIMENT

To engineer the Xe related defect, we performed ion implantation into two separate diamonds – a type IIA chemical vapour deposition (CVD) diamond with 1 ppm nitrogen (Element Six) and an electronic grade CVD diamond with <1 ppb nitrogen (see figure 1(a)). Using these two samples enables to correlate between the PL intensity of the Xe emitters and the nitrogen dopant concentration in diamond. Ion implantation was performed at room temperature using a National Electrostatics Corporation 4UH ion accelerator with 500 keV $^{129}$Xe$^+$ ions to a total dose of $1\times10^{12}$, $1\times10^{11}$, $1\times10^{10}$, $1\times10^9$ Xe ions/cm$^2$ (figure 1(a)) The dose rate was $3 \times 10^{12}$ - $5 \times 10^{12}$ /cm$^2$/s depending on the respective quadrant. Irradiations were performed at 7° off the [100] direction to minimize channelling effects. These doses have been used previously for various dopants to study dose dependent luminescence and identification of single quantum emitters. The depth profile of implanted $^{129}$Xe atoms under experimental conditions was calculated using the SRIM software package[18] (density = 3.515 g / cm$^3$, displacement energy = 40 eV) (Figure 1b). A single Gaussian fit to the implantation profile calculated by SRIM produces a peak at a depth of 105 nm and a half width at half maximum (HWHM) of 26 nm. After implantation, all samples were annealed at 1400°C for one hour under a pure nitrogen atmosphere.

Optical measurements were performed at room temperature using a standard confocal microscope with a high numerical aperture objective (0.7 NA). Green (532 nm) and red (633 nm) lasers were used for off resonant excitation, and a pulsed 532 nm (40 MHz, 100 ps pulses) laser for lifetime measurements. For the resonant excitation (PLE) measurements a Ti:sapphire laser (~ sub 5-MHz linewidth) that is tuned to the high energy ZPL (~ 791 nm) was used. The PLE scan experiment was performed by scanning a piezo etalon within a mode hop-free range of ~50 GHz and detecting the off resonant phonons or the 2$^{nd}$ ZPL (~ 812 nm). A wave meter was used to monitor the wavelength of the excitation laser, and provide feedback to stabilize the laser at a particular wavelength.

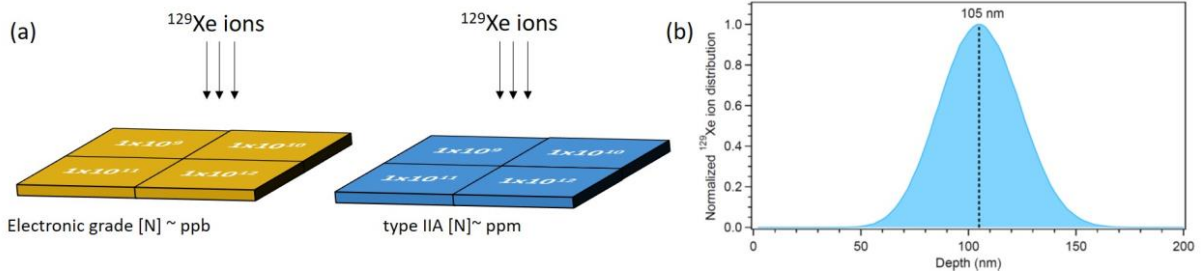

Figure 1. (a) 500 keV $^{129}$Xe ions were implanted into two separate diamonds – electronic grade with ~ ppb nitrogen concentration and type IIA with ~ppm nitrogen concentration. (b) Implantation depth profile of 500 keV $^{129}$Xe ions into single crystal diamond calculated using the SRIM software package. A single Gaussian fit to the implantation profile produces a peak depth of 105 nm and a FWHM of 26 nm.

## III. Results and Discussions

First, we studied the formation of Xe related color centers in the implanted diamonds. Figure 2(a) shows luminescence spectra from the electronic grade sample using green (532 nm) excitation at room temperature. The Xe related ZPL doublet is clearly observed in the higher implantation dose region ($1\times10^{12}$ Xe ions/cm$^2$) and is still observed at the lower dose of $1\times10^{11}$ Xe ions/cm$^2$. However, no signature of the Xe related defects was observed from the lowest implantation doses, even with high resolution confocal scanning. Notably, in the 1×10$^{10}$ Xe ions/cm$^2$ region, single nitrogen vacancy (NV) emitters were found (as expected from the electronic grade diamond after ion irradiation and annealing). This implies that the quantum efficiency of the Xe related color center is at least an order of magnitude lower than that of the NV when pumped at room temperature using a 532 nm laser.

To gain increased information about the defects, we compared the fabricated emitters in the electronic grade and type IIA diamond samples. Figure 1(b) shows a comparison of the emitters excited by the 633 nm laser. We note that this comparison is not possible using 532 nm excitation since emission spectra of the type IIA sample under this condition is dominated by NV fluorescence (not shown). Interestingly, the intensity of the emission in the nitrogen rich sample is an order of magnitude higher than in the ultra-pure material. This suggests that the Xe center is negatively charged, since the presence of nitrogen is likely to provide excess electrons for defect formation. A similar effect has also been reported for the negatively charged silicon vacancy (SiV) defect in diamond, which displays an increased intensity in the presence of nitrogen. Despite the higher efficiency in nitrogen rich diamond, we were not able to identify single Xe related centers in the regions of lowest implantation doses even in the type IIA sample.

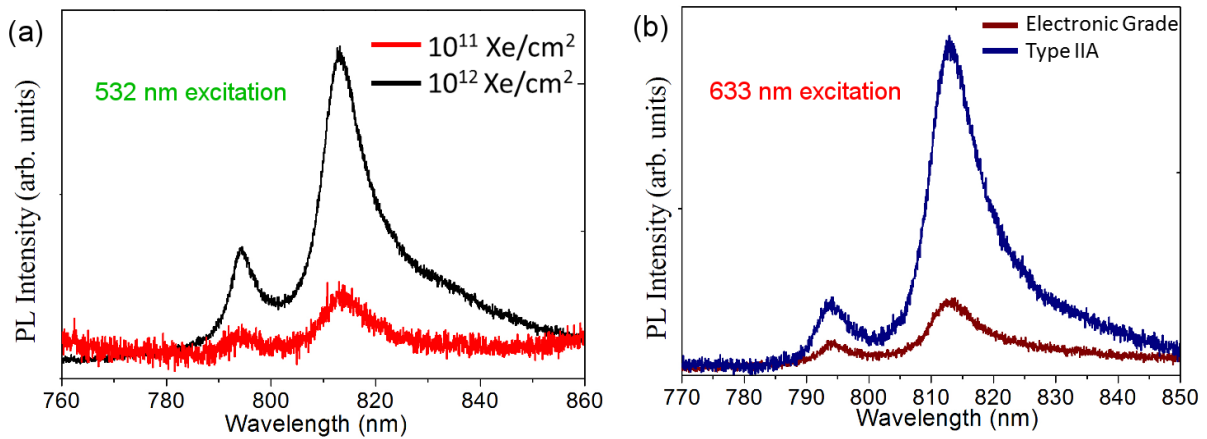

*Figure 2. (a) Room temperature PL spectra from Xe related defects in electronic grade diamond recorded from 1×10$^{12}$ (black curve) and 1×10$^{11}$ Xe ions/cm$^2$ (red curve) implantation areas, respectively. Two ZPLs at 794 nm and 811 nm are clearly visible. (b) Comparison of luminescence intensity of the Xe color centers from the electronic grade and the type IIA sample. Both measurements recorded from the 1×10$^{12}$ Xe ions/cm$^2$ area under the same laser power and collection conditions.*

Next, we measured the fluorescence lifetime of an ensemble of Xe emitters (Figure 3). A very short excited state lifetime of ~ 0.73 ns and ~ 0.77 ns was measured for the Xe related color centers in type IIA and the ultrapure diamond, respectively. The laser decay is shown in comparison to elucidate the system response (~ 0.43 ns). The extremely short lifetime further supports the presence of non-radiative relaxation pathways from the excited state via phonon transitions. The observed lifetime is nevertheless comparable to the fluorescence lifetime of the SiV defect, which despite being very dim, can indeed be isolated as a single site after fabrication by ion implantation.

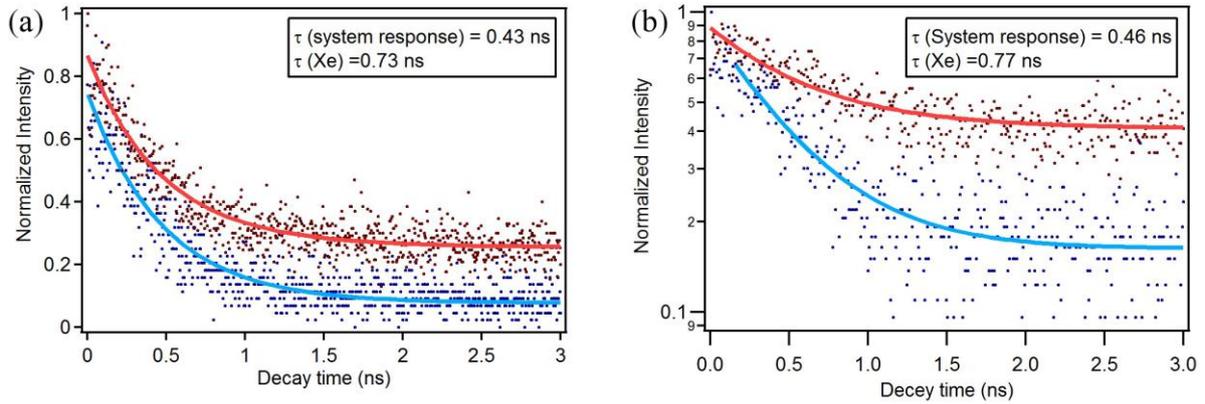

*Figure 3. Excited state lifetime of the Xe related color centers in (a) type IIA diamond, and (b) ultrapure material. The measurement was recorded from the 1×10$^{12}$ Xe ions/cm$^2$ and the Xe related lifetime is below 1 ns.*

While isolation of single Xe based emitters was not successful using off resonant excitation at room temperature, the optical properties of Xe defects at cryogenic temperatures can still be investigated. Figure 4a shows an off resonant excitation of an ensemble of Xe emitters. Both the high energy ZPL (~ 794 nm) and the low energy ZPL (~ 811 nm) are visible at low temperature. The FWHM of both ZPLs were reduced to ~ 0.88 nm and ~ 0.81 nm separately. The inset of Figure 4a is the polarization measurement of 811 nm ZPL emission that confirms the probed Xe centers are fully polarized, and the small probed ensemble is fully aligned along the same direction. As shown in figure 4b, the Xe related center has two excited states. Due to the fast phonon relaxation passage (indicated by the thick dash down arrow), population in the upper excited state is lower than that of the lower ground state. Hence the 811 nm ZPLs is stronger than the 794 nm ZPL at low temperature, suggesting the thermalization of the 794 nm lie.

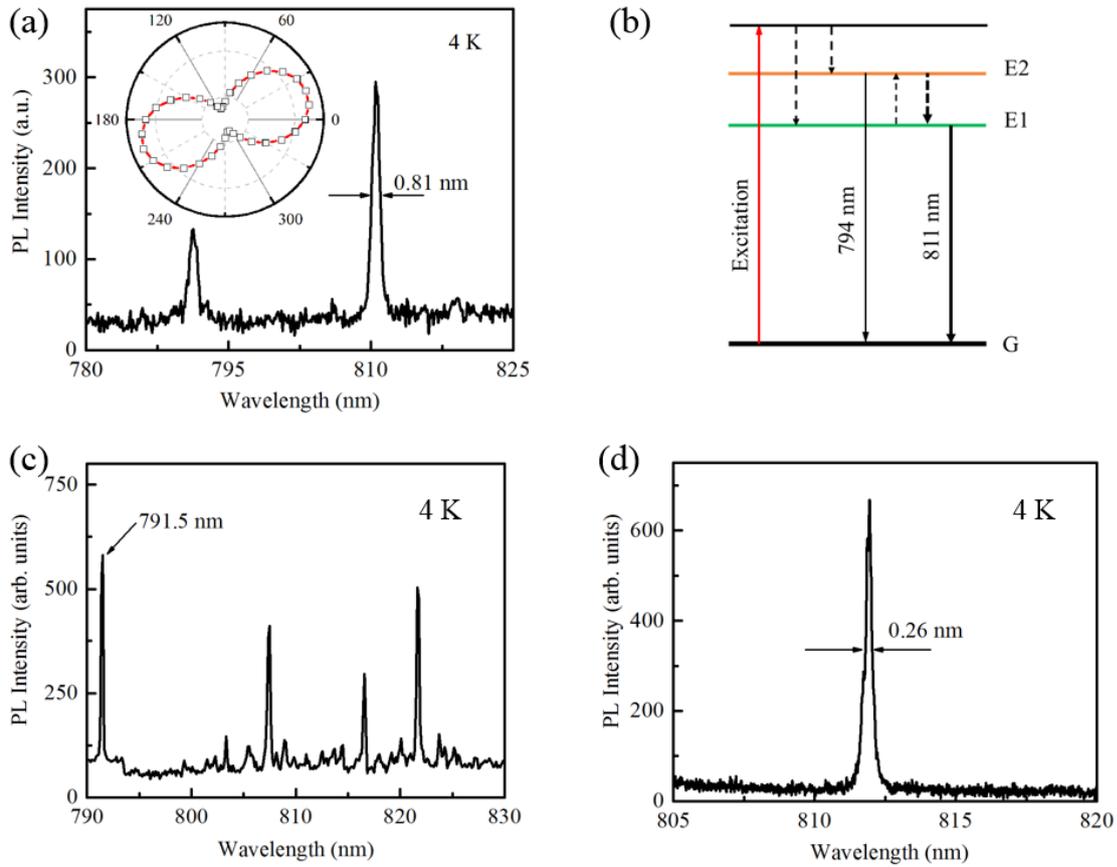

*Figure 4. (a) Off resonant excitation (745nm 2.2 mW) at 4K, showing the sub nm Xe related ZPL lines at 791.3 nm and 810.6 nm separately. Inset is the polarization of the 810.6 nm ZPL. The 791.3 nm line was blocked by a long pass filter. (b) Schematic Energy level of Xe related color center in diamond. The Xe related center is excited off resonance to a virtual absorption state, and relaxes back to E1 and E2 through photon relaxation passage. The thick down dash arrow between state E2 and E1 indicate a fast decay phonon relaxation passage, while the phonon relax passage from state E1 to E2 is weaker (thin up dash arrow). The thick black down arrow indicate a stronger 811 ZPL transition. (c) Resonant excitation of the Xe color centers using the higher energy ZPL at 791.5 and monitoring the lower energy ZPLs. The observed lines are all spectrometer limited (~ 120GHz) and are likely correspond to single Xe related color centers. (d) Another ensemble of Xe related centers demonstrate single and narrow 811 nm ZPL when resonant excited with 280 μW, 971.5 nm laser light.*

To further demonstrate resonant excitation, we excite the high energy ZPL at 791.5 nm and collect the second ZPL and the associated phonon side bands as shown in Figure 4(c). The observed spectra reveals several narrow lines, that are spectrometer limited (~ 120 GHz) and a weak phonon side band. As shown in Figure 4(d), we manage to find a point where only one 811 nm ZPL was recorded. We tune the excitation wavelength across 791.5 nm, and found when excitation wavelength is near resonant with the 791.5 nm ZPL transition, the collected 811 nm ZPL signal is the strongest. We attribute this to the strong correlation between the high energy level (E2) and low energy level (E1) of the Xe related centers. This result is promising for further studies of resonant excitation of Xe related centers and their use for nanophotonics applications.

Finally, we measured the ensemble PL response under a high magnetic field by exciting resonantly at the higher energy ZPL of ~ 793.8 nm and measuring the luminescence of the lower energy ZPL at ~ 811.7 nm under a varying magnetic field from 0 to 9 T. The excitation power is fixed at 50 μW to avoid apparent heating. The external magnetic field is aligned along the optical axis of the microscope and corresponds to the [001] axis of the diamond. We set our spectrometer at its highest resolution (1800 g/mm) to observe the narrow 811.7 nm ZPL at different magnetic field. Under zero magnetic field, the narrow 811.7 nm ZPL was observed to have two separate components as depicted in Figure 5(a). When the magnetic field was varied from 0 T for 9 T, the ZPL peaks do not shift in frequency (wavelength) and we always observe a constant linewidth for each individual peaks as shown in Figure 5(b) and 5(c). We conclude there is no split within the resolution of our measurement when the Xe related centers was subjected to different magnetic fields in a wide range from 0 T to 9 T. As show in Figure 5(d), the only change is the signal of peak 2 drops when magnetic field was increased.

The PL response of Xe related center under a high magnetic field indicates that both ground and excited states have the same Landé factors (g values) and therefore a similar spin state in both the ground and excited state for both ZPLs[19, 20]. A weak spin–orbit coupling can also induce transitions to different spin states that will be undetectable using the current setup, therefore, at this stage we cannot confirm the absolute spin state of the ground and excited state. We note, however, that for the case of a S=1 system, like the NV defect, no splitting in the ZPL is expected due to similar g values in the ground and excited states, while for S=1/2 systems, such as the SiV defect, the ZPL lines do shift under applied magnetic fields[20]. This is promising for the future application of the Xe color centers in spin related applications and observation of optically detected magnetic resonance. Furthermore, it is interesting that the observed emission intensity of the lower energy ZPL is reducing with an increased magnetic field (Figure 5c). This indicates a strong spin dependent transition. Under an increasing magnetic field, a splitting in the ground or the excited state occurs that is strongly coupled to a non-radiative or a long-lived state that results in a reduced overall emission intensity.

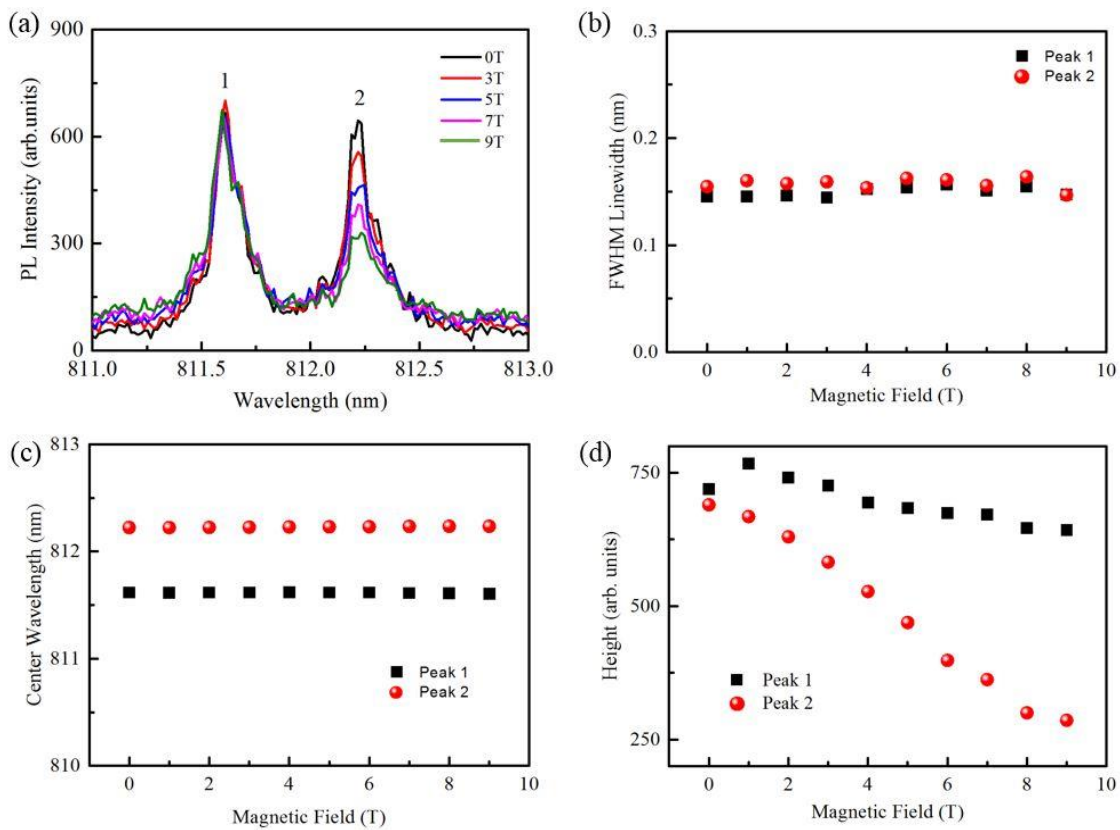

*Figure 5. (a) The ZPLs spectrum of Xe related center under different magnetic field when the center is resonant excited. The excitation wavelength is 793.765 nm and excitation power is 50 μW. (b) and*

*(c)The linewidths and center wavelength of each individual peak do not change when magnetic field was varied from 0 T to 9 T. (d) The signal instensity of both peaks drops when magnetic field was increased.*

## IV. Conclusions

To summarize, we presented a detailed optical characterization of an ensemble of Xe related optical defects in diamond. We showed that the emitters can be efficiently created using ion implantation, possess a very short excited state lifetime and exhibit brighter emission in a nitrogen doped diamond which is indicative of a negative charge state. Furthermore, we demonstrated resonant excitation and performed luminescence studies under strong magnetic field that indicate similar g values in both ground and excited states. The intensity of the low energy ZPL depends on the magnetic field, likely via coupling to a long lived metastable state. Further studies into the spin properties of the Xe centers are therefore warranted. While room temperature single Xe related emitters were not observed in this study, this may be possible in the future by increasing the collection efficiency using nanoscale pillars[21] solid immersion lenses[22] or etched into the diamond or using nanodiamonds coupled with specially designed dielectric resonators[23]. Moreover, the promising resonant excitation experiments along with the strong ZPL, indicate that the Xe color center in diamond is a promising system to couple to photonic resonators such as microdisks or photonic crystals.

## V. Acknowledgments

A portion of this work was performed under the auspices of the U.S. Department of Energy by Lawrence Livermore National Laboratory under Contract No. DE-AC52-07NA27344. Financial support from the Australian Research Council (DE130100592), FEI Company, and the Asian Office of Aerospace Research and Development grant FA2386-15-1-4044 are gratefully acknowledged. WBG acknowledges the support from Singapore 2015 NRF fellowship grant (NRF-NRFF2015-03) and its Competitive Research Programme (CRP Award No. NRF-CRP14-2014-02).